# Design, Development and Testing of a Conformal 60 GHz Active Repeater for High Energy Physics Applications

Imran Aziz[1,2*], Yasin Alekajbaf[1], Dragos Dancila[1], Kristiaan Pelckmans[1], Abhinav Jain[3], Richard Brenner[1]

[1] Department of Physics and Astronomy, Uppsala University, Uppsala, Sweden. * imran.aziz@physics.uu.se
[2] Department of Electrical Engineering, Mirpur University of Science and Technology (MUST), Mirpur AJK, Pakistan.
[3] Department of Electrical Engineering, Uppsala University, Sweden.

*Abstract*—The Wireless Allowing Data and Power Transmission (WADAPT) proposal was formed to study the feasibility of wireless technologies in HEP experiments. A strong motivation for using wireless data transmission at the ATLAS detector is the absence of wires and connectors to reduce the passive material. However, the tracking layers are almost hermetic, acting as a Faraday cage, that doesn't allow propagation between the layers. For radial readout of the detector data through the layers, we have developed an active repeater board, which is passed through a 2-3 mm wide slit between the modules on the tracking layers. The repeater is also advantageous in building topological radial networks for neuromorphic tracking. The active repeater board consists of an RX antenna, an amplifier, and a TX antenna, and is tested on a mockup in a way that the RX antenna will be on the inner side of a module, and the TX antenna will be on the outer side of the same module, as the 10-mil thick conformal board is passed through the small slit. Transmission through the tracking layers using the repeater has been demonstrated with two horn antennas, a signal generator, and a spectrum analyzer. For 20 cm distance between the horn antenna and the repeater board, a receive level of -19.5 dBm was achieved. In comparison, with the same setup but with the amplifier turned off, the receive level was ~-46 dBm. The results show a significant milestone towards the implementation of 60 GHz links for detector data readout.

## I. Introduction

The European Organization for Nuclear Research (CERN) is home to the world's largest particle physics experiments, including the ATLAS detector [1] [2], which plays a critical role in probing fundamental particles and interactions. ATLAS generates vast amounts of data during high-energy particle collisions, requiring efficient, high-speed communication systems for data acquisition and readout. Currently, this data transmission relies on an extensive network of optical cables to handle the high bandwidth requirements. However, the use of optical cables within the detector presents several challenges. These include space constraints, mechanical vulnerability, abundance of passive material, and the complexity of maintaining a large number of physical connections. Additionally, as the detector continues to evolve with the High-Luminosity Large Hadron Collider (HL-LHC) upgrade, there is an increasing demand for scalable, flexible, and less intrusive solutions to handle the projected increase in data flow.

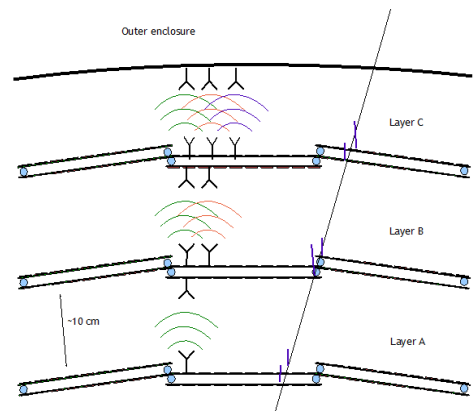

Fig. 1. *Tracking layers at the ATLAS detector as CERN, showing a proposal of radial data readout through wireless links.*

The Wireless Allowing Data and Power Transmission (WADAPT) [2] proposal was formed to study the feasibility of wireless technologies in High Energy Physics (HEP) experiments at CERN. To address the limitations of optical read out system, we propose a wireless communication network operating at millimeter-wave frequencies, specifically at 60 GHz, as an alternative to the current optical infrastructure. This transition from optical to wireless technology offers potential benefits such as reducing cable complexity, improving flexibility in system layout, and enhancing the overall robustness of the data transmission network within the detector. However, the proposed solution involves some challenges, e.g., the tracking layers at the detector, schematic shown in Fig. 1, are almost hermetic, acting as a Faraday cage, that allows multiple links in the volume without severe crosstalk, but doesn't allow propagating the signal between the layers. For propagation between the layers, we have designed and developed an active repeater board, which is passed through a couple of millimeters (mm) wide slit between the modules on each layer. The repeater is also advantageous in building topological radial networks for neuromorphic tracking, as the current readout system employs optical cable that can only take out the signal axially, resulting in higher latency.

## II. Repeater Design

### A. Microstrip Patch Antenna

A single layer Rogers 5880 (epsr = 2.2) with 10 mil height is used to design a microstrip patch antenna. The material has been chosen keeping in view its conformal properties for the desired application. The antenna is designed in Ansys Electronics. Fig. 2. shows the top view of the antenna, designed using the inset-fed technique. A lumped port is used as an input. Part (b) of the figure presents the E- and H-plane radiation pattern, while a maximum realized gain of 8.6 dBi is achieved at 60 GHz. The S11 for the antenna is presented in part (c) of the figure, having less than -10 dB impedance bandwidth from 58 GHz to 61.5 GHz, with the lowest value at around 60 GHz.

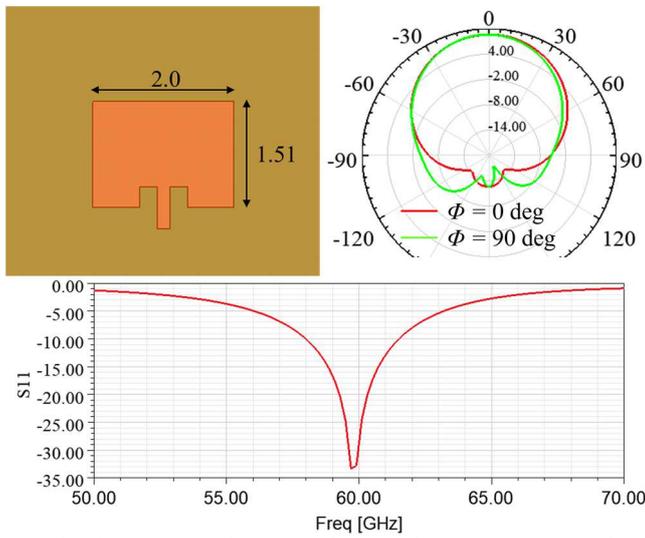

Fig. 2. *Antenna design with its radiation pattern and S11 (simulations)*

### B. Amplifier

A low noise amplifier (LNA) from Analog Devices (HMC-ALH382) [3] is used to design the repeater structure for coverage beyond a single detector layer. The LNA has 21 dB gain, noise figure 3.8 dB, P1dB 12 dBm, and a die size of 1.55 x 0.73 x 0.1 mm.

### C. Repeater

The repeater circuit employs a receiving antenna (RX ANT), an amplifier (AMP), and then the transmitting antenna (TX ANT). The RX ANT receives the 60 GHz signal from the lower layer, gets amplified by the AMP, and re-transmitted by the TX ANT on the other side of the detector layer.

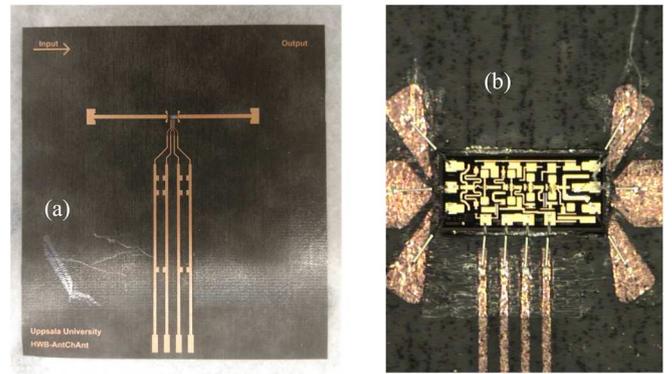

Fig. 3. *Fabricated active repeater board (a) amplifier is wire-bonded on the board (b).*

## III. Fabrication and Characterization

Fig. 3 shows the fabricated structure of the repeater PCB. On the left is the RX ANT, AMP in the center, and the TX ANT on the right. The 4-lines coming down from the center are for the DC-biasing of the amplifier. Part (b) of the figure shows the wire-bonded amplifier die, which can be understood as a zoom-in view of the center of the repeater PCB. There are two quarter wave long open-ended stubs on each side of the 50-ohm line, intended to make a short to connect with the ground pad of the die. The repeater board is tested on a mockup shown in Fig. 4 (a, b, & c). The board is fixed on the mockup in a way that the receiving antenna will be, e.g., on the inner side of Layer B (Fig. 1), and the transmitting antenna will be on the outer side of the same layer, as the 254-um thick conformal board is passed through the small slit between modules.

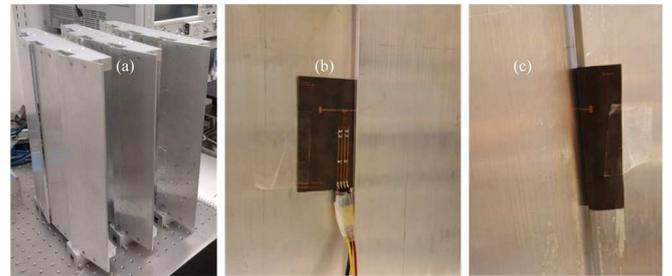

Fig. 4. *Mockup to test the conformal repeater (a) receiving side of the board is mounted on the inner side of the mockup, where the wires coming down are for the DC-biasing (b) patch antenna to re-transmit the amplified signal on other side of the mockup (c).*

The repeater structure has been characterized and transmission though the tracking layers have been demonstrated using the schematic setup shown in Fig. 5. 2 x horn antennas are used to test the conformal repeater PCB. The TX horn antenna is getting the 60 GHz signal from a signal generator. The RX ANT on the repeater board receives this signal, gets amplified by the LNA, and re-transmitted by the TX ANT on the other side of the layer. This signal is received by the other horn antenna, followed by a V-band 14[th]- harmonic mixer that gets its local oscillator (LO) signal from another signal generator at 4.276 GHz. The intermediate frequency (IF) is then read at the Signal Hound spectrum analyzer at 136 MHz (IF = RF – 14 x LO). The frequencies have been corrected in the measurement results shown in Fig 6. For a 20 cm distance between the horn antenna and the repeater board, the RX level is -19.5 dBm, while for the same setup, when the amplifier the OFF, the RX level is -45.8 dBm. The results represent a significant milestone towards the implementation of 60 GHz wireless links for the detector data readout.

In the next step, we are working on integrating the ST 60 GHz transceiver chip with the amplifier and antenna to design a complete 60 GHz wireless link. The ST transceiver chip shall be fed with 5 Gbps data through an RF-SoC (radio frequency Silicon on chip).

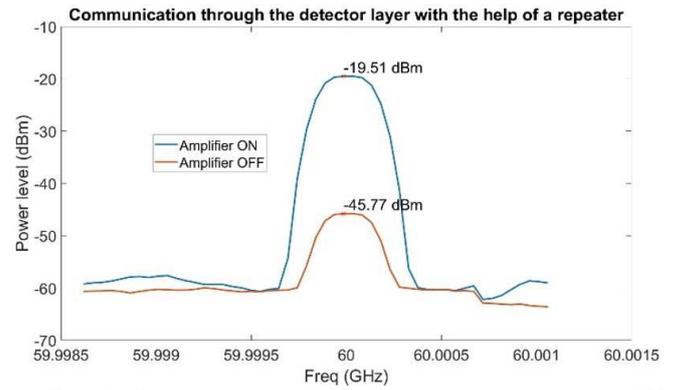

Fig. 6. *Corrected power levels when the amplifier is ON (blue) and when the amplifier is OFF (orange).*

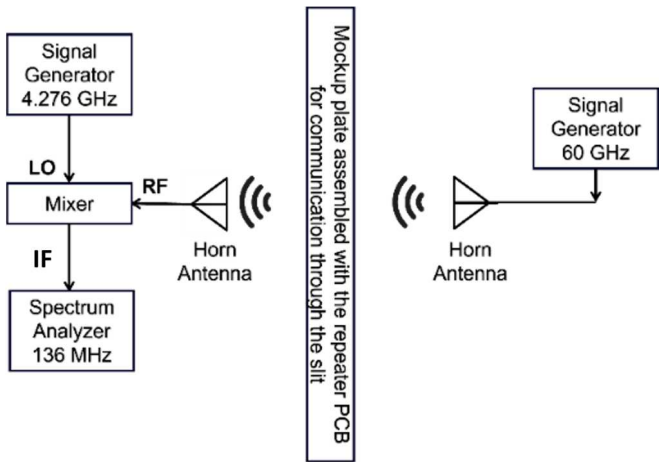

Fig. 5. *Measurement setup schematic for propagation through the tracking layer with the help of active repeater board. The RX patch antenna on the repeater board, mounted on the mockup, receives the signals from the horn antenna on the right, and the TX patch antenna re-transmits the signal to the horn antenna on the left.*